\documentclass[amsmath,floatfix,twocolumn,superscriptaddress]{revtex4-1}
\usepackage{amssymb,amsmath}
\usepackage{graphicx,array}
\usepackage{dcolumn}
\usepackage{psfrag}
\usepackage{bm}
\usepackage{color}

\begin{document}
\title{Strain engineering of oxide thin films for photocatalytic applications}

\author{Zhao Liu}
\affiliation{School of Materials Science and Engineering, UNSW Sydney, NSW 2052, Australia} 
\affiliation{Sino-French Institute of Nuclear Engineering and Technology, Sun Yat-sen University, Zhuhai 519082, China} 

\author{Joel Shenoy}
\affiliation{School of Materials Science and Engineering, UNSW Sydney, NSW 2052, Australia}

\author{Cesar Men\'{e}ndez}
\affiliation{School of Materials Science and Engineering, UNSW Sydney, NSW 2052, Australia}

\author{Judy N. Hart}
\affiliation{School of Materials Science and Engineering, UNSW Sydney, NSW 2052, Australia} 

\author{Charles C. Sorrell}
\affiliation{School of Materials Science and Engineering, UNSW Sydney, NSW 2052, Australia} 

\author{Claudio Cazorla}
\affiliation{School of Materials Science and Engineering, UNSW Sydney, NSW 2052, Australia}

\begin{abstract}
Photocatalytic materials are pivotal for the implementation of disruptive clean energy 
applications such as conversion of H$_{2}$O and CO$_{2}$ into fuels and chemicals driven 
by solar energy. However, efficient and cost-effective materials able to catalyze the chemical 
reactions of interest when exposed to visible light are scarce due to the stringent 
electronic conditions that they must satisfy. Chemical and nanostructuring approaches are 
capable of improving the catalytic performance of known photoactive compounds however the 
complexity of the synthesized nanomaterials and sophistication of the employed methods 
make systematic design of photocatalysts difficult. Here, we show by means of first-principles 
simulation methods that application of biaxial stress, $\eta$, on semiconductor oxide thin films 
can modify their optoelectronic and catalytic properties in a significant and predictable manner. 
In particular, we show that upon moderate tensile strains CeO$_{2}$ and TiO$_{2}$ thin films 
become suitable materials for photocatalytic conversion of H$_{2}$O into H$_{2}$ and CO$_{2}$ into 
CH$_{4}$ under sunlight. The band gap shifts induced by $\eta$ are reproduced qualitatively 
by a simple analytical model that depends only on structural and dielectric susceptibility 
changes. Thus, epitaxial strain represents a promising route for methodical screening and 
rational design of photocatalytic materials. 
\end{abstract}
\maketitle

\section{Introduction}
\label{sec:intro}
The adverse effects of burning fossil fuels and the growing concentration of CO$_{2}$ in the atmosphere 
are motivating intense research efforts towards large-scale production of clean fuels and conversion 
of carbon dioxide into useful chemicals. In this context, generation of H$_{2}$ from water and 
reduction of CO$_{2}$ into methane (CH$_{4}$) and other valuable substances by using energy from 
sunlight represent two very promising sustainable approaches \cite{fajrina19,xie16}. 

Sunlight-induced dissociation of H--O and C--O bonds in water environment involves the use of photocatalytic 
materials that should fulfill quite stringent electronic requirements. For instance, the band gap ($E_{g}$) 
of photocatalytic materials must be below $\sim 3$~eV in order to absorb solar radiation within the visible 
spectral range. At the same time, a semiconductor able to catalyze the synthesis of H$_{2}$ in water or 
the reduction of CO$_{2}$ into CH$_{4}$ should possess a conduction-band edge higher in energy than the 
corresponding redox potential ($-4.4$ and $-4.6$~eV, respectively, relative to the vacuum level) and 
the valence-band edge lower than the H$_{2}$O oxidation potential ($-5.6$~eV) \cite{xie16,shenoy19}. 
Furthermore, active photocatalysis requires efficient separation of the photogenerated charge carriers 
(low exciton binding energy) and their rapid transportation to the reaction sites (long electron-hole 
recombination time) \cite{park16}.

Cerium and titanium dioxide, CeO$_{2}$ and TiO$_{2}$, are two extensively investigated catalyst materials 
that present high structural stability, commercial availability, and low toxicity. Examples of applications 
in which CeO$_{2}$ and TiO$_{2}$ are exploited include fuel and solar cells, water purification, 
corrosion-resistant coatings, therapeutic agents, and gas sensors, to cite just a few \cite{rahimi16,castano15,montini16}.
Nevertheless, the band gap of both catalysts are larger than $3.0$~eV, which severely restricts their absorbance 
of sunlight (to only $\approx 3$\% of the solar irradiation that reaches the earth's surface \cite{kowalsa08}). 
Chemical and nanostructuring strategies have been employed successfully to reduce the band gaps of CeO$_{2}$ 
and TiO$_{2}$ and thus improve their photocatalytic activity under sunlight \cite{asahi01,mitsudome11}. Nevertheless, 
the usual complexity associated with the energy landscapes of nanomaterials and the required synthesis methods 
make it difficult to identify what key parameters improve photocatalytic efficiency irrespective of the material 
\cite{park16}. As a consequence, progress in ``photocatalysis by design'', which is different from just ranking 
photocatalyst materials by their performance (``black box screening''), remains limited \cite{takanabe17}. In 
addition to such design constraints, precious co-catalysts like Pt and Au typically are employed for improving 
photocatalytic performance, which is not suitable for practical applications \cite{xie16,kowalsa08}. 

In this article, we show by means of first-principles simulations based on density functional theory that 
biaxial strain, $\eta$ (which can be achieved in practice by growing thin films on substrates presenting 
a lattice parameter mismatch through, for example, pulsed laser deposition techniques \cite{hu18,heo17}), can 
be exploited to tune the optoelectronic and photocatalytic properties of some binary oxides in a substantial and 
controlled manner. Specifically, we predict that under feasible tensile strains of $\approx +2$ and $+3$\% 
\cite{gopal17,benson17} CeO$_{2}$ and TiO$_{2}$ become suitable materials for conversion of H$_{2}$O into 
H$_{2}$ and of CO$_{2}$ into CH$_{4}$ under sunlight. Such potential enhancements in photocatalytic activity 
result from sizeable band gap reductions ($\sim 10$\%) and correct positioning of the valence and conduction-band 
edges under $\eta$. Meanwhile, the effects of epitaxial strain on the band gap of ZnO, another well-known 
semiconductor photocatalyst \cite{choi18}, are found to be only marginal ($|\Delta E_{g}| / E_{g} \sim 1$\%). 
We present quantitative and physically intuitive arguments that explain the origins of such irregular $\eta$-driven 
effects on $E_{g}$ in terms of dielectric susceptibility and metal-oxygen bond length changes.

\begin{figure*}[t]
\centerline{
\includegraphics[width=1.00\linewidth]{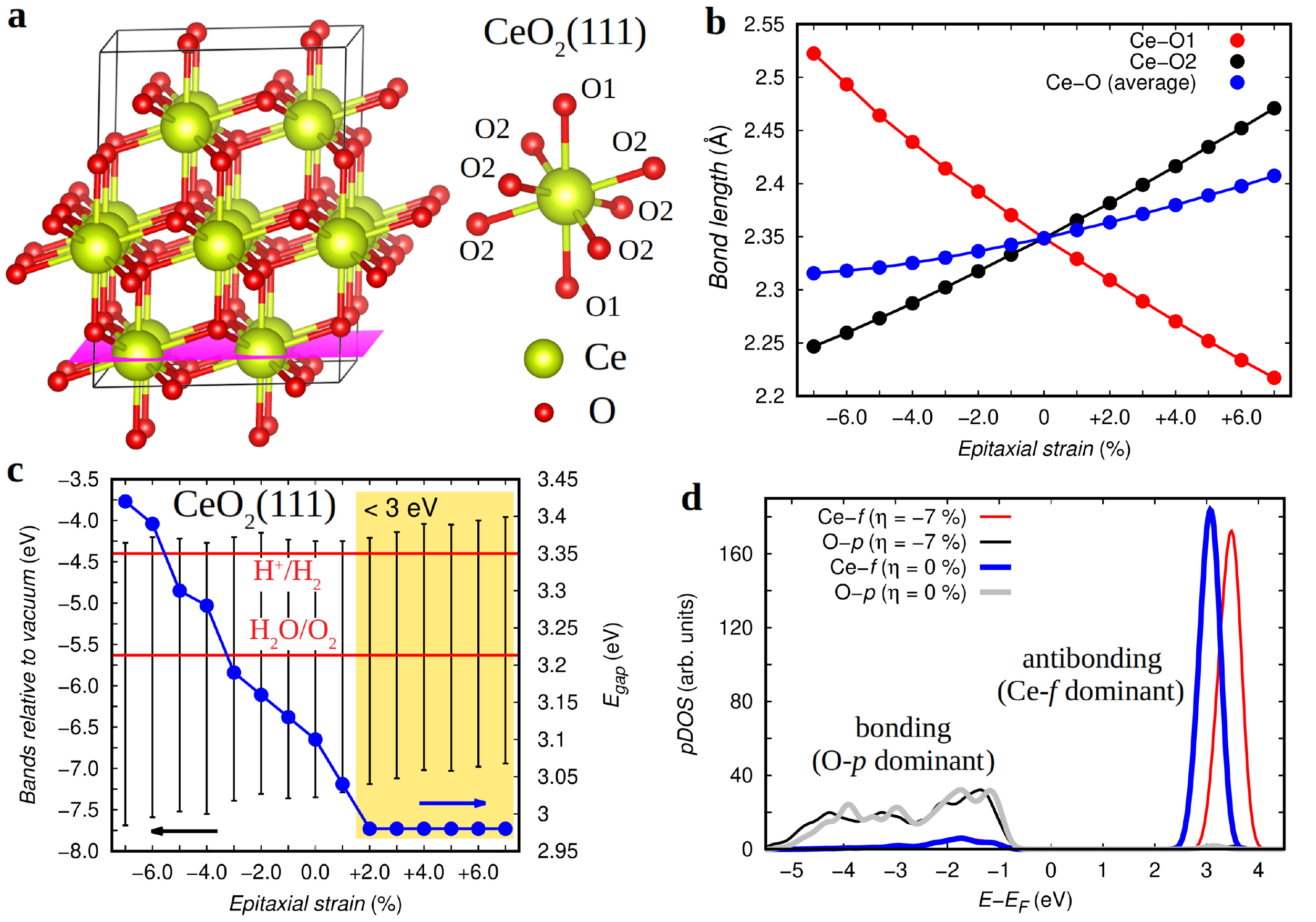}}
\caption{Strain engineering of fluorite CeO$_{2}$~(111). {\bf a} Atomic structure of the analyzed thin film system. 
	The plane in which biaxial strain is applied is indicated in pink. The local oxygen environment of 
	the Ce atoms is shown. {\bf b} Variation of different Ce-O bond lengths as a function of biaxial strain.
	{\bf c} Band gap and band alignment changes induced by epitaxial strain. The region in which the system band
	gap is lower than $3$~eV is highlighted in yellow. The redox potentials of interest for water splitting 
	are indicated with red horizontal lines. {\bf d} Partial density of states calculated around the Fermi energy 
	level at different $\eta$ conditions.}
\label{fig1}
\end{figure*}

It is worth noting that recent experimental and theoretical works have already proposed epitaxial strain as a means 
for tuning $E_{g}$ in some oxide materials such as TiO$_{2}$ \cite{yin10,kelaidis18}, ZnO \cite{choi18}, SnO$_{2}$ 
\cite{zhou14}, and CdO \cite{yan12}. However, a clear and general understanding of how biaxial strain affects the 
photocatalytic performance of binary oxides is still missing. For instance, the relative $E_{g}$ variations induced 
by tensile biaxial strain are negative in some materials (band gap decreases in TiO$_{2}$ and SnO$_{2}$ \cite{yin10,zhou14}) 
whereas positive in others (band gap increases in ZnO \cite{choi18}). Likewise, the $E_{g}$ changes driven by compressive 
biaxial strain are positive in some materials (TiO$_{2}$ and SnO$_{2}$ \cite{yin10,zhou14}) whereas negative or almost 
null in others (ZnO \cite{choi18} and CdO \cite{yan12}). These results indicate that the causes of $\eta$-induced band 
gap shifts cannot be traced down uniquely to simple structural changes \cite{yin10,zhou14,wagner02} since those changes 
are quite monotonous under biaxial strain regardless of the material. Moreover, the influence of $\eta$ on the band 
alignments of binary oxides has been neglected in previous studies despite their potential impact on the photocatalytic 
activity. Hence, the present theoretical work identifies the key factors that drive $\eta$-induced band gap changes and 
fills the existing knowledge gaps. The results presented next show that strain engineering, either used on its own or as 
a complement to other existing approaches, can be a powerful tool for rational design and systematic improvement of 
photocatalytic materials.

\section{First-principles computational methods}
\label{sec:methods}
First-principles calculations based on density functional theory (DFT) \cite{cazorla15a,cazorla12b,cazorla17a} are 
performed to simulate and analyze the influence of biaxial strain ($\eta$) on several representative binary oxide 
photocatalysts. We use the PBEsol functional \cite{pbesol} as is implemented in the VASP software package \cite{vasp}. 
A ``Hubbard--$U$'' scheme \cite{hubbard} with $U = 3$~eV is employed for a better treatment of the localized Ce $4f$, 
Ti $3d$, and Zn $3d$ electronic orbitals. We use the ``projector augmented wave'' method to represent the ionic cores 
\cite{paw} by considering the following electrons as valence: Ce $4f$, $5d$, $6s$, and $4d$; Gd $4f$, $5d$, and $6s$; 
Ti $3d$, $4s$, $3p$, and $3s$; Zn $3d$ and $4s$; and O $2s$ and $2p$. Wave functions are represented in a plane-wave 
basis truncated at $650$~eV. For integrations within the Brillouin zone we employ Monkhorst-Pack {\bf k}--point grids 
\cite{kpoint} with a density equivalent to that of $16 \times 16 \times 16$ in the fluorite CeO$_{2}$ unit cell. 
Strained-bulk geometry relaxations are performed with a conjugate-gradient algorithm that allows for volume variations 
while imposing the structural constraints defining thin films ($|a|=|b|$ and $\alpha = 90^{\circ}$) 
\cite{cazorla15,cazorla17,cazorla17b,cazorla18}. Periodic boundary conditions are applied along the three directions 
defined by the lattice vectors, so that possible surface effects are completely avoided in the simulations. The relaxations 
are halted when the forces acting on the atoms fall below $0.01$~eV$\cdot$\AA$^{-1}$. By using these technical parameters 
we obtain zero-temperature energies that are converged to within $0.5$~meV per formula unit. Biaxial strain conditions 
are simulated at $\Delta \eta = 1$\% intervals. In order to estimate accurate band gaps and band alignments, we employ 
the hybrid HSE06 exchange-correlation functional \cite{hse06} to perform single-point calculations on the equilibrium 
geometries determined at the PBEsol$+U$ level \cite{shenoy19}. The generation of non-stoichiometric and Gd-doped fluorite 
CeO$_{2}$ thin films are explained in the Supplementary Methods along with some details of their energy and structural 
properties. 

To calculate accurate band alignments we follow the work done by Moses and co-workers on binary semiconductors 
\cite{moses11}. Briefly, both bulk and slab calculations are performed from which the alignment of the electrostatic 
potential within the semiconductor material can be obtained relative to the vacuum level. From the slab calculations, 
the difference between the average electrostatic potential within the semiconductor material and in vacuum is obtained. 
From the bulk calculations, the band structure shifts relative to the average electrostatic potential are determined. 
These calculations are performed at each $\eta$ point and involve the estimation of macroscopic and planar average 
potentials (Supplementary Methods). The planar potential is computed by averaging potential values within a well defined 
plane (for instance, perpendicular to the surface of the slab), and the macroscopic potential is obtained by taking 
averages of the planar potential over distances of one unit cell along the chosen direction \cite{resta88,cazorla12}.
The slab systems should be thick enough to ensure that the electron density in the center of the slab is practically 
equal to that in the bulk material. We have found that $1.2$--$1.8$~nm thick oxide slabs accompanied by similarly 
large portions of vacuum provide sufficiently well converged results for the electrostatic potentials. Further 
technical details of our band alignment calculations are provided in the Supplementary Methods.

\section{Results and Discussion}
\label{sec:results}
The changes induced by biaxial strain, $\eta$, on the structural, electronic, and photocatalytic properties of
CeO$_{2}$, TiO$_{2}$, and ZnO as calculated with first-principles methods based on density functional theory
(DFT, Sec.~\ref{sec:methods} and Supplementary Methods) are presented first. Both compressive ($\eta < 0$) 
and tensile ($\eta > 0$) biaxial strains ranging from zero up to a maximum absolute value of $7$\% have been 
considered in the simulations. These values are comparable in magnitude to the $\eta$'s achieved experimentally 
in the same materials \cite{gopal17,benson17,choi18}. At the end of this section, we introduce a simple analytical 
model that depends only on structural and dielectric susceptibility changes and reproduces qualitatively, and 
helps in understanding, the general band gap trends induced by $\eta$.

\subsection{Fluorite CeO$_{2}$~(111)}
\label{subsec:ceo2-111} 
At room temperature, bulk ceria (CeO$_{2}$) presents a cubic phase known as fluorite (space group $Fm\overline{3}m$) 
in which the Ce ions form a face centered cubic sublattice and the oxygens a simple cubic. Ceria thin films commonly 
exhibit three high-symmetry orientations $\{111\}$, $\{011\}$, and $\{001\}$. In this section, we consider the $\{111\}$ 
case; results for the $\{001\}$ geometry will be presented in the next section. 

In the fluorite structure each Ce ion is surrounded by eight equidistant oxygens. Upon application of (111) biaxial 
strain, some crystal symmetries are broken (space group changes to $R\overline{3}m$) and two characteristic metal-oxygen 
bond lengths emerge, Ce--O1 and Ce--O2, which are six- and two-fold degenerate, respectively (Fig.\ref{fig1}a). The Ce--O1 
bonds are oriented perpendicular to the (111) plane, in which the strain is applied, and thus under compressive (tensile) 
biaxial strain they are stretched (reduced). Conversely, the Ce--O2 bonds, which are mostly contained within the (111) plane, 
are shortened (stretched) under compressive (tensile) biaxial strain (Fig.\ref{fig1}b). Due to the higher degeneracy 
of the Ce--O2 bonds, the average metal-oxygen neighbouring length decreases under compressive biaxial strain and increases 
under tensile $\eta$ (Fig.\ref{fig1}b). For instance, at the maximum simulated compressive (tensile) biaxial strain the 
average Ce--O distance is reduced (elongated) by $1.4$~($2.5$)\% of the unstrained value. 

The sizable structural changes induced by $\eta$ suggest the possibility of finding similarly large 
variations in the band gap ($E_{g}$) of (111)--oriented CeO$_{2}$ thin films \cite{yin10,zhou14}. In fact, 
as shown in Fig.\ref{fig1}c, this turns out to be the case. Specifically, $E_{g}$ increases practically 
linearly under compressive biaxial strain, reaching a maximum value of $3.4$~eV at $\eta = -7$\%. According 
to our DFT calculations the band gap of unstrained CeO$_{2}$ is $3.1$~eV, which compares very well 
with the experimental value of $3.2$~eV \cite{goubin04}, hence a maximum relative $E_{g}$ increase of $10$\% 
is achieved. Under tensile biaxial strain, $E_{g}$ also varies significantly although not in a regular manner. 
The band gap first decreases to below $3.0$~eV at $\eta \approx +2$\% but beyond that strain point its value remains 
practically constant (Fig.\ref{fig1}c). As a result, a maximum $\eta$-driven $E_{g}$ reduction of $4$\% is 
obtained. It is worth noting that the largest $E_{g}$ variation is achieved under compressive $\eta$ whereas 
the largest average Ce--O length change is achieved under tensile strain. This observation suggests that, at 
least for (111)--oriented CeO$_{2}$ thin films, the band gap shifts induced by $\eta$ cannot be explained exclusively 
in terms of the accompanying structural changes.  

Figure~\ref{fig1}c also shows the band alignments of (111)--oriented CeO$_{2}$ thin films, that is, the energy level 
of the valence-band (VB) and conduction-band (CB) edges as a function of $\eta$. In the absence of any strain, our 
calculations predict a VB edge located at $-7.3$~eV and CB at $-4.2$~eV with respect to the vacuum level, which are 
in fairly good agreement with the available experimental data ($E_{\rm VB}^{\rm expt} = -6.9$~eV and $E_{\rm CB}^{\rm expt} 
= -4.1$~eV \cite{wen18}). For photocatalytic water-splitting purposes, the electronic band structure of unstrained 
CeO$_{2}$~(111) presents two important limitations. First, the corresponding $E_{g}$ is too large for absorption of
visible light, and second, the position of the VB (CB) edge is too far below (too close to) the water oxidation potential 
of $-5.6$~eV (the proton reduction potential of $-4.4$~eV) \cite{xie16,shenoy19}. Interestingly, biaxial strain can be 
used to partly overcome both of those limitations. Under tensile strain, the energy of the VB edge increases steadily 
and becomes equal to $-6.9$~eV at $\eta = +7$\%, $0.4$~eV closer to the water oxidation potential than for unstrained
CeO$_{2}$. At the same strain, the CB edge reaches a maximum value of $-4.0$~eV and the band gap becomes smaller than 
$3$~eV. Thus, under $\eta > 0$ sunlight can be absorbed more efficiently and the VB and CB edges are situated more 
appropriately for water splitting (that is, $1.3$~eV below and $0.4$~eV above the corresponding redox potentials). 
Meanwhile, under compressive strain the energy of the VB edge decreases steadily and the CB level remains more or less 
constant. As a consequence, the band gap of CeO$_{2}$~(111) increases almost linearly with increasing strain magnitude
in the $\eta < 0$ region. We note that the band gap of unstrained bulk ceria is indirect and remains so in the investigated 
$\eta$ interval (Supplementary Figure~1).

The influence of biaxial strain on the VB and CB edges of CeO$_{2}$~(111) (and thus on the band gap, defined as 
$E_{g} \equiv E_{\rm CB} - E_{\rm VB}$) can be understood in terms of the concomitant electronic and structural 
changes. In bulk ceria, the top of the VB is mostly composed of oxygen $2p$ orbitals that form a bonding state
with Ce $4f$ orbitals, while the bottom of the CB is mostly composed of cerium $4f$ orbitals that form an antibonding 
state with O $2p$ orbitals (Fig.\ref{fig1}d). Compressive (tensile) strain reduces (increases) the average Ce--O bond 
length, which energetically favors (frustrates) the bonding state. Consequently, the energy of the VB edge decreases 
under $\eta < 0$ and increases under $\eta > 0$. On the other hand, the bottom of the CB is found to be quite insensitive 
to compressive strains (Fig.\ref{fig1}c). When the anion-cation bond lengths are reduced, the kinetic energy associated 
with the antibonding state typically increases (since it is proportional to $k^{2}$, where $k$ represents the reciprocal 
lattice vector in the extended Brillouin zone) \cite{wei99}; that increase in kinetic energy would bring the CB higher 
in energy. However, the $p$--$f$ level repulsion appears to diminish slightly under compressive strain owing to the 
increased delocalization of the Ce $4f$ orbitals (see density of states peaks above the Fermi energy level in Fig.\ref{fig1}d,
where the Ce $4f$ states extend over a wider energy range for $\eta < 0$). These two effects tend to oppose each 
other thus leaving the CB edge unaffected by compressive strain. Under tensile strain, the energy of the antibonding state 
eventually rises due to a significant increase in the localization of the Ce $4f$ orbitals (Supplementary Figure~2) which 
enhances the $p$--$f$ level repulsion and overcomes the accompanying decrease in kinetic energy.  

\begin{figure*}[t]
\centerline{
\includegraphics[width=1.00\linewidth]{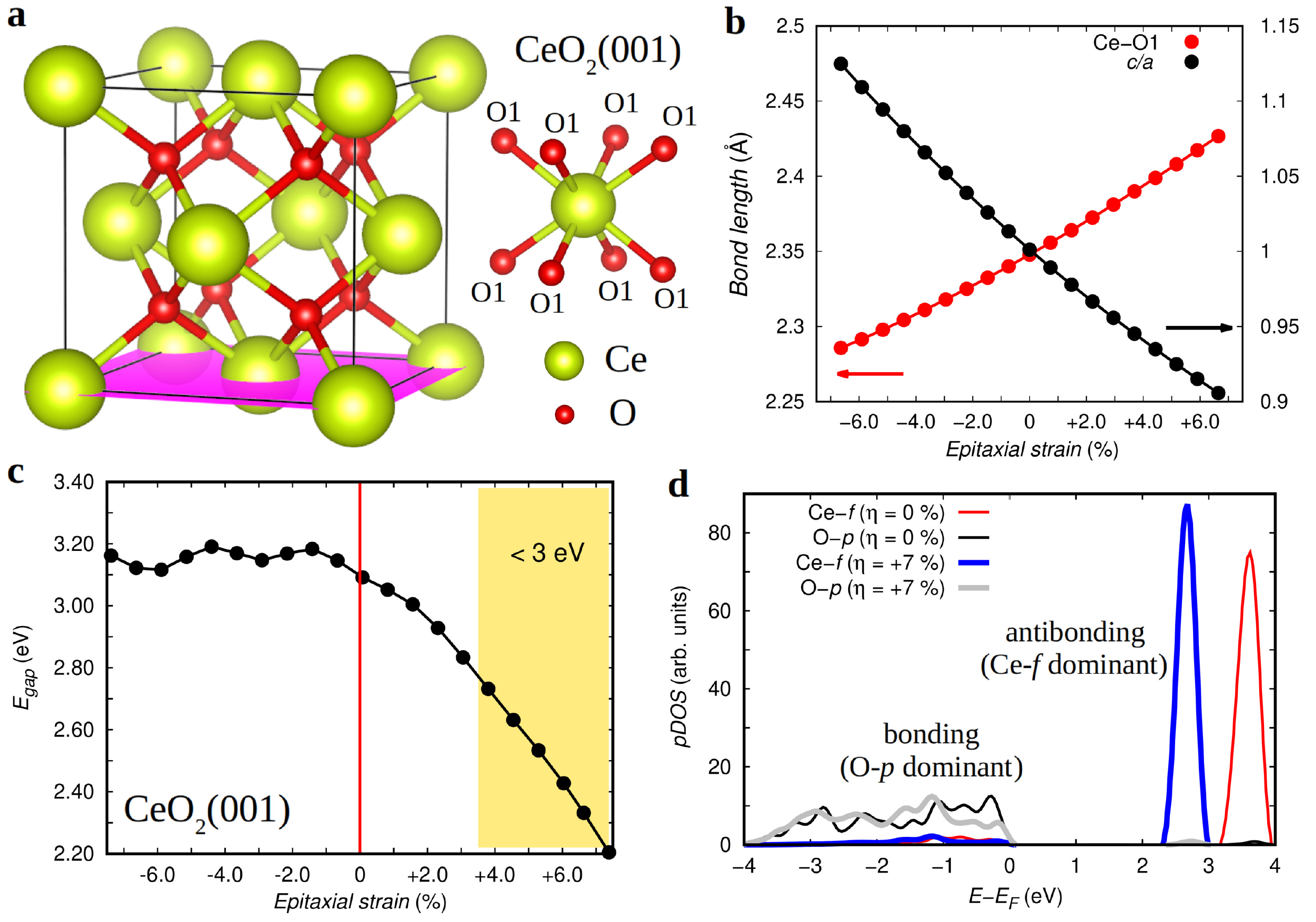}}
\caption{Strain engineering of fluorite CeO$_{2}$~(001). {\bf a} Atomic structure of the analyzed thin film system.
        The plane in which biaxial strain is applied is indicated in pink. The local oxygen environment of the
        Ce atoms is shown. {\bf b} Variation of different structural parameters as a function of biaxial strain.
        {\bf c} Band gap changes induced by epitaxial strain. The region in which the system band gap is lower than 
	$3$~eV is highlighted in yellow. {\bf d} Partial density of states calculated around the Fermi energy level
        at different $\eta$ conditions.}
\label{fig2}
\end{figure*}

\subsection{Fluorite CeO$_{2}$~(001)}
\label{subsec:ceo2-001}
Upon application of (001) biaxial strain, the degeneracy of the eight equidistant oxygens surrounding each Ce ion 
(characteristic distance Ce--O1 in Fig.~\ref{fig2}a) is not lifted, but the symmetry of the crystal changes from 
cubic to tetragonal (space group $I4/mmm$) due to contraction or elongation of the out-of-plane $c$ axis relative to 
the two in-plane lattice vectors $a = b$ (Fig.\ref{fig2}b). The changes driven by (001) strain on $E_{g}$ are noticeably 
different from those found in the (111) case (Figs.\ref{fig2}c and \ref{fig1}c). In particular, the band gap now is
hardly affected by compressive strain but is significantly reduced under tensile strain. For instance, at $\eta = +7$\% 
the band gap is $2.2$~eV, $\approx 30$\% lower than the value obtained at equilibrium conditions; at $\eta \approx +3$\%, 
$E_{g}$ is already smaller than $3.0$~eV. Plots of the electronic energy bands indicate that the band gap remains 
indirect regardless of $\eta$ (Supplementary Figure~3), as for the CeO$_{2}$~(111) thin films.
 
The influence of $\eta$ on the band edges of (001)--oriented CeO$_{2}$ thin films can also be rationalized in terms of 
the accompanying structural and electronic changes. In this case, we have not explicitly calculated the band alignments 
however, in analogy to the (111) case, we assume that the energy of the VB edge decreases under $\eta < 0$ (increases 
under $\eta > 0$) due to enhancement (frustration) of the $p$--$f$ bonding interactions induced by shortening (elongation) 
of the Ce--O1 bonds. Based on this assumption and there being no effect of compressive strain on the band gap, it is expected 
that the energy of the CB edge decreases under compressive strain due to increased delocalization of the unoccupied Ce $4f$ 
orbitals (Supplementary Figure~4), which weakens the $p$--$f$ antibonding interactions and counteracts the concurrent increase 
in kinetic energy \cite{wei99}. Hence, both the VB and CB edges decrease under $\eta < 0$ producing no net change in the band 
gap. Under tensile strain, however, it can be inferred from the decrease in the band gap that the bottom of the conduction band 
remains more or less constant with respect to the vacuum level, possibly due to a compensation effect between the increased 
localization of the Ce $4f$ orbitals (see density of states peaks above the Fermi energy level in Fig.\ref{fig2}d, where the 
Ce $4f$ states extends over a narrower energy range for $\eta > 0$), which tends to bring the CB higher in energy, and the 
reduction in kinetic energy, which tends to lower it. 

\begin{figure*}[t]
\centerline{
\includegraphics[width=1.00\linewidth]{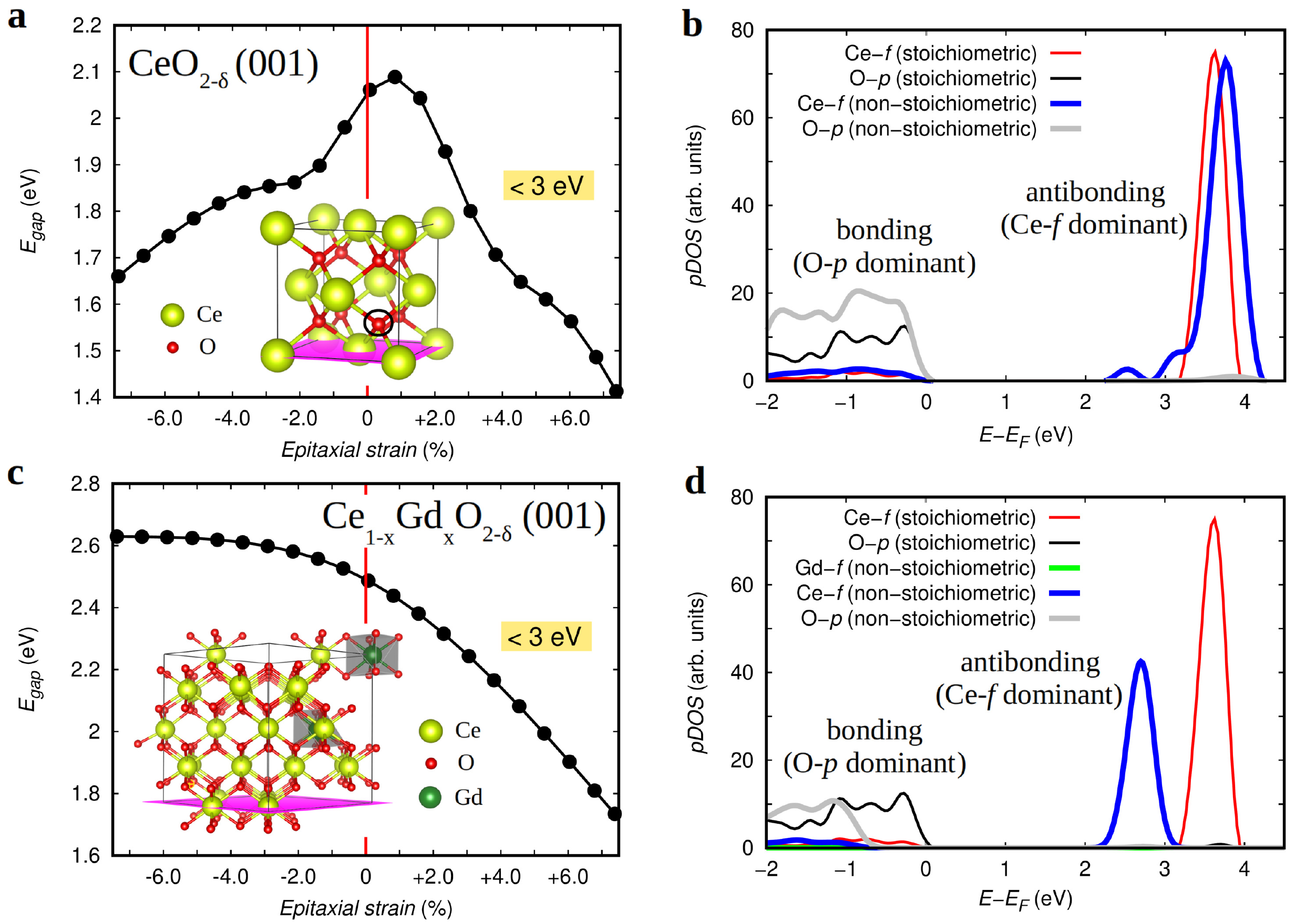}}
\caption{Strain engineering of fluorite CeO$_{2-\delta}$ and Ce$_{1-x}$Gd$_{x}$O$_{2-\delta}$~(001). 
	{\bf a},{\bf c} Atomic structure and band gap of CeO$_{2-\delta}$ and Ce$_{1-x}$Gd$_{x}$O$_{2-\delta}$~(001), 
	respectively, as a function of biaxial strain. The planes in which biaxial strain is applied are 
	indicated in pink. The position of the oxygen vacancy created in CeO$_{2-\delta}$~(001) is indicated 
	with a black circle. {\bf b},{\bf d} Partial density of states calculated around the Fermi energy level in  
	stoichiometric and non-stoichiometric CeO$_{2-\delta}$ and Ce$_{1-x}$Gd$_{x}$O$_{2-\delta}$~(001), respectively.}
\label{fig3}
\end{figure*}

It has been demonstrated thus far that biaxial strain can have a significant impact on the optoelectronic     
properties of stoichiometric CeO$_{2}$. It is reasonable to ask then whether similar control of functionality  
can be achieved in non-stoichiometric and metal-doped ceria thin films with $\eta \neq 0$, which can be prepared 
through advanced synthesis techniques \cite{mofarah19,korobko12}. The results shown in Fig.\ref{fig3} indicate that 
this is indeed the case. Our simulations of non-stoichiometric ceria, CeO$_{2-\delta}$, considering an arbitrary 
but representative $\delta$ of $0.125$ \cite{mofarah19} (Sec.~\ref{sec:methods} and Supplementary Methods), show 
that both positive and negative $\eta$'s can be used to reduce substantially $E_{g}$ (by $\sim 10$\% of the value 
obtained at zero strain, which is $2.1$~eV), with tensile strain being particularly effective (Fig.\ref{fig3}a). 
Band gaps of $1.7$ and $1.4$~eV are obtained respectively at the highest compressive and tensile strain values 
considered in this study. These sizable $E_{g}$ reductions result from the combined action of oxygen vacancies, 
which are known to reduce the neighbouring Ce$^{4+}$ ions and lower the CB edge due to the appearance of new 
unoccupied $4f$ states (Fig.\ref{fig3}b) \cite{mofarah19}, and biaxial strain. 

Likewise, in Ce$_{1-2x}$Gd$_{2x}$O$_{2-x}$~(001) thin films with an arbitrary but representative composition of 
$x = 0.03$ \cite{korobko12}, our DFT simulations (Sec.~\ref{sec:methods} and Supplementary Methods) indicate that 
tensile strain is also capable of reducing $E_{g}$ considerably (Fig.\ref{fig3}c). For instance, at $\eta = +7$\% 
the band gap of Gd-doped ceria is $1.7$~eV, which is approximately $30$\% smaller than the value estimated at zero 
strain ($2.5$~eV). In Gd-doped ceria, the presence of Gd$^{3+}$ ions prevents the reduction of Ce$^{4+}$ ions 
surrounding the oxygen vacancies (in contrast to what occurs in CeO$_{2-\delta}$). As a consequence, the relative
$E_{g}$ variation induced by $\eta$ in Ce$_{1-2x}$Gd$_{2x}$O$_{2-x}$~(001) is very similar to that found in 
the undoped stoichiometric system (Fig.\ref{fig2}c). The high delocalization of the Ce $4f$ orbitals forming the 
bottom of the CB in Ce$_{1-2x}$Gd$_{2x}$O$_{2-x}$~(001) (Fig.\ref{fig3}d), however, leads to a noticeable reduction 
in the band gap as compared to undoped CeO$_{2}$~(001), which renders a practically constant $E_{g}$ difference of 
$\approx 0.5$~eV between the two systems across the entire range of $\eta$ values. 

\begin{figure*}[t]
\centerline{
\includegraphics[width=1.00\linewidth]{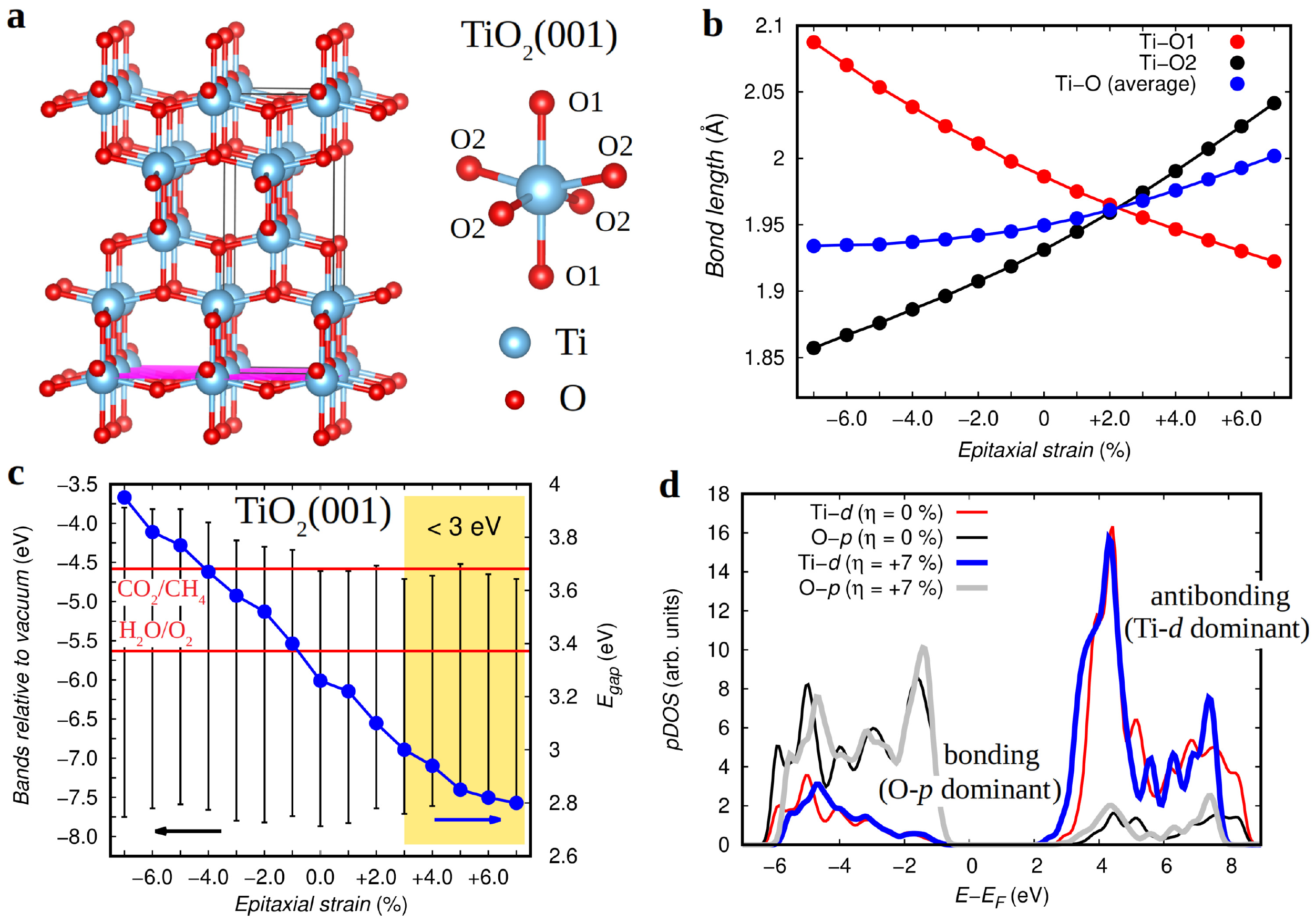}}
\caption{Strain engineering of anatase TiO$_{2}$~(001). {\bf a} Atomic structure of the analyzed thin film system.
        The plane in which biaxial strain is applied is indicated in pink. The local oxygen environment of the
        Ti atoms is shown. {\bf b} Variation of different Ti-O bond lengths as a function of biaxial strain.
        {\bf c} Band gap and band alignment changes induced by epitaxial strain. The region in which the system 
	band gap is lower than $3$~eV is highlighted in yellow. The redox potentials of interest are indicated
        with red horizontal lines. {\bf d} Partial density of states calculated around the Fermi energy level
        at different $\eta$ conditions.}
\label{fig4}
\end{figure*}

\subsection{Anatase TiO$_{2}$~(001)}
\label{subsec:tio2-001}
Bulk TiO$_{2}$ is generally found in the rutile, anatase, or brookite phase. The rutile and anatase polymorphs 
find use in major industrial applications while brookite is of little technological relevance due to its difficult 
synthesis \cite{yamada12}. Rutile possesses tetragonal symmetry (space group $P4_{2}/mmm$) and a band gap of $\approx 
3$~eV, and is the energetically most favorable polymorph at room temperature. Anatase also presents tetragonal 
symmetry (space group $I4_{1}/amd$) and its band gap is $3.2$~eV \cite{yamada12}. Although the $E_{g}$ of anatase 
is larger than that of rutile, the former phase typically exhibits superior photocatalytic activity owing to better 
positioning of the band edges and longer electron-hole recombination times \cite{batzill11,choi94}. Hence it is of 
particular interest to investigate the influence of $\eta$ on the band gap and band alignments of anatase, to see 
whether it is possible to further improve the photocatalytic performance of TiO$_{2}$.   

In anatase TiO$_{2}$, each Ti ion is surrounded by six neighbouring oxygens and forms two characteristic bonds 
with them, Ti--O1 and Ti--O2, which are respectively two- and four-fold degenerate (Fig.\ref{fig4}a). When 
biaxially strained in the (001) plane, the Ti--O1 bonds, which are oriented out-of-plane, become elongated 
under compressive strain and shortened under tensile strain (Fig.\ref{fig4}b). Conversely, the Ti--O2 bonds, 
which are oriented at only a small angle to the (001) plane, are reduced under $\eta < 0$ and stretched 
under $\eta > 0$ (Fig.\ref{fig4}b). Consequently, as in most oxide semiconductors \cite{cazorla17,bousquet10}, 
the average metal-oxygen bond length decreases under compressive strain (by $0.8$\% of the equilibrium value at 
$\eta = -7$\%) and increases under tensile strain (by $2.7$\% at $\eta = +7$\%).

\begin{figure*}[t]
\centerline{
\includegraphics[width=1.00\linewidth]{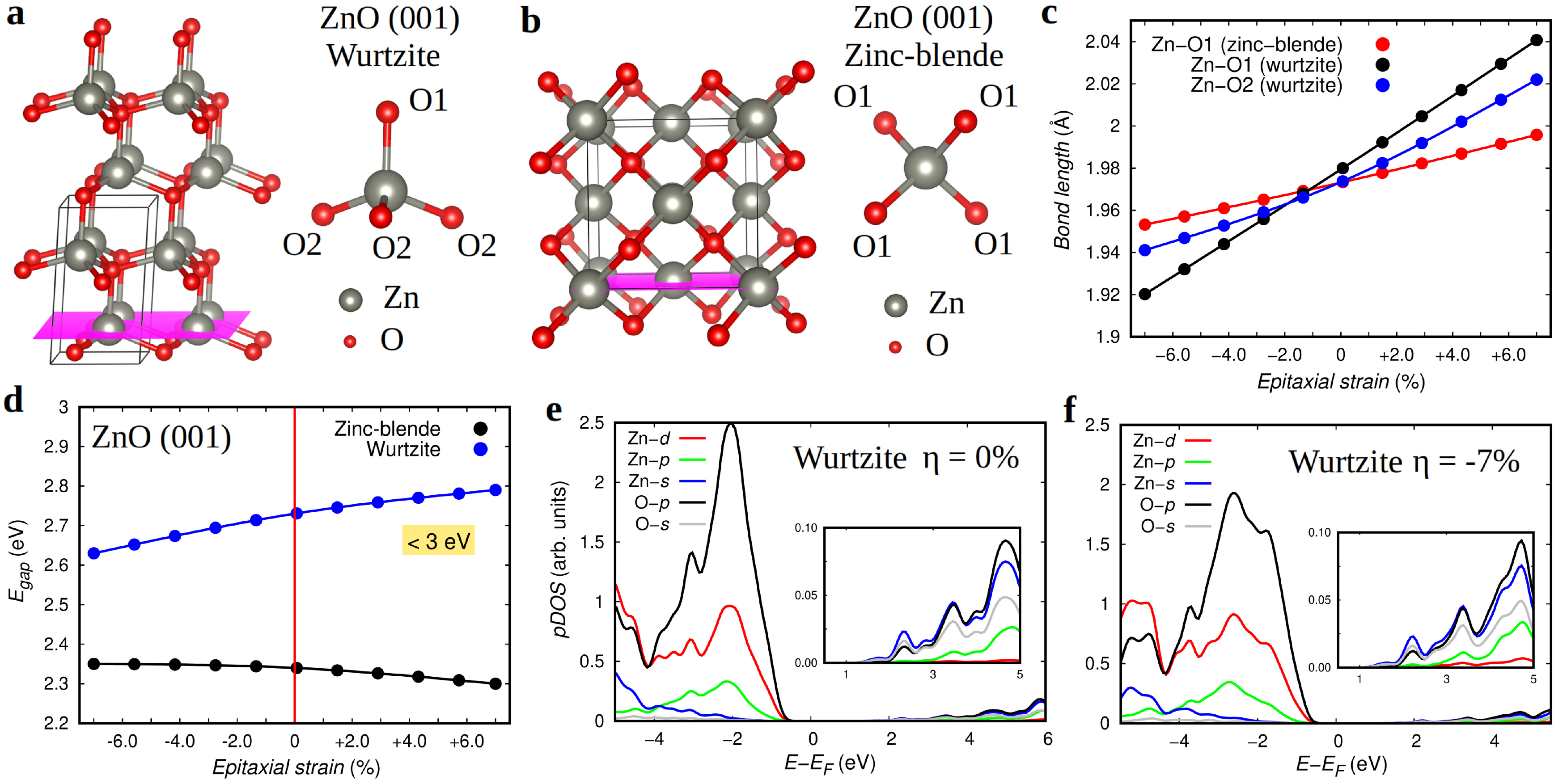}}
\caption{Strain engineering of wurtzite and zinc-blende ZnO~(001). Atomic structure of the analyzed 
	{\bf a} wurtzite thin film and {\bf b} zinc-blende thin film. The planes in which 
	biaxial strain is applied are indicated in pink. The local oxygen environment of the Zn atoms are 
	shown. {\bf c} Variation of different Zn-O bond lengths as a function of biaxial strain for the two 
	crystal structures. {\bf d} Band gap changes induced by epitaxial strain. {\bf e} Partial density of 
	states calculated around the Fermi energy level at equilibrium and {\bf f} the maximum compressive 
	biaxial stress considered in this study. The insets show the states at the conduction band minima 
	in more detail.}
\label{fig5}
\end{figure*}

Concerning the electronic properties, at zero biaxial strain we estimate $E_{g} = 3.3$~eV, which is in good 
agreement with the experimental value of $3.2$~eV \cite{yamada12}. Under compressive strain, the band gap increases
almost linearly, achieving a maximum value of $4.0$~eV at $\eta = -7$\% (Fig.\ref{fig4}c). Under tensile strain,
however, the band gap decreases steadily and becomes smaller than $3$~eV at $\eta \approx +3$\%. A minimum band
gap of $2.8$~eV is estimated at the largest tensile distortion considered in this study (Fig.\ref{fig4}c). We 
note that $E_{g}$ remains indirect regardless of $\eta$ (Supplementary Figure~5). 

Such a regular $E_{g}$ variation driven by biaxial strain across the entire range of strain values is in 
contrast to what was found for both the (111) and (001) orientations of CeO$_{2}$, and can be explained in 
terms of the concurrent VB and CB edge shifts (Fig.\ref{fig4}c). In bulk anatase, the top of the VB is 
principally composed of oxygen $2p$ orbitals that form a bonding state with Ti $3d$ orbitals while the bottom 
of the CB is mostly composed of Ti $3d$ orbitals that form an antibonding state with O $2p$ orbitals (Fig.\ref{fig4}d).
Compressive strain has little effect on the position of the VB edge, due to a compensation effect between 
enhancement of bonding interactions (which tends to lower the VB) and increase in kinetic energy (which tends
to increase the VB), whereas tensile strain tends to bring it higher in energy, owing to frustration of the 
bonding state (which dominates over the decrease in kinetic energy). Meanwhile, the position of the CB edge 
moves significantly higher in energy under compressive strain, due to a dominant increase in the kinetic energy 
(since the localization of the unoccupied Ti $3d$ orbitals appears to decrease slightly at $\eta < 0$, Supplementary 
Figure~6). Under tensile strain, the CB edge remains more or less constant owing to a small increase in the 
localization of the unoccupied Ti $3d$ orbitals at $\eta > 0$ (Fig.\ref{fig4}d) that is counterbalanced by a 
decrease in the kinetic energy. We note that, in analogy to CeO$_{2}$ thim films, the $E_{g}$ shifts induced 
by $\eta$ do not appear to be directly or exclusively correlated with the accompanying structural changes (i.e., 
the band gap variation is almost linear across the entire range of $\eta$ values, Fig.\ref{fig4}c, whereas the 
structural fluctuations are most prominent under tensile strain, Fig.\ref{fig4}b).

The band alignments in anatase TiO$_{2}$~(001) are greatly affected by biaxial strain (Fig.\ref{fig4}c). In 
the absence of planar stress, our calculations render a VB edge at $-7.9$~eV and CB at $-4.6$~eV relative 
to the vacuum level, which are in reasonable agreement with the available experimental data ($E_{\rm VB}^{\rm expt} = 
-7.6$~eV and $E_{\rm CB}^{\rm expt} = -4.4$~eV \cite{batzill11}). Under tensile strains of $> +3$\%, the band gap  
becomes smaller than $3$~eV, the corresponding VB edge gets closer to the H$_{2}$O oxidation potential (for 
instance, at $\eta = +4$\%, $E_{\rm VB}$ is $-7.6$~eV), and the CB edge remains more or less constant around 
the value $-4.6$~eV. At these biaxial strain conditions, the energy of the CB edge is around the reduction potential 
of CO$_{2}$ (Fig.\ref{fig4}c), hence tensilely strained anatase is predicted to be a suitable photocatalyst for driving 
the conversion of carbon dioxide into methane (CH$_{4}$) in aqueous environment under visible light \cite{xie16}. 
Given the proximity of the CO$_{2}$ and water reduction potentials, it is likely that biaxially strained anatase 
TiO$_{2}$ ($\eta > 0$) is also a suitable material for driving the production of hydrogen fuel from H$_{2}$O under 
visible light.

\subsection{Zinc-blende and wurtzite ZnO~(001)}
\label{subsec:zno-001}
ZnO thin films are used in a wide variety of optoelectronic applications due to the relative abundance of their 
elements and direct band gap of $\approx 3.3$~eV \cite{ozgur05}. The two common ZnO polymorphs are wurtzite 
(hexagonal symmetry, space group $P6_{3}mc$) and zinc-blende (cubic symmetry, space group $F\overline{4}3m$),
which are shown in Figs.\ref{fig5}a-b. In wurtzite ZnO, each metal ion is coordinated to four neighboring oxygens 
and forms two characteristic metal-oxygen bond lengths, Zn--O1 and Zn--O2 (Fig.\ref{fig5}a), which are single 
and three-fold degenerate, respectively. The evolution of the Zn--O1 and Zn--O2 distances as induced by $\eta$
in the (001) plane are shown in Fig.\ref{fig5}c. The Zn--O1 bond length is oriented out-of-plane and displays 
an anomalous behavior in the sense that it contracts under compressive biaxial strain and expands under tensile 
biaxial strain. Such an anomalous behavior, which is not observed in the Zn--O2 case (Fig.\ref{fig5}c), probably 
is related to the negative thermal expansion observed in bulk ZnO \cite{wang13}. On average, however, the mechanical 
behavior of wurtzite ZnO thin films is normal, namely, volume contraction at $\eta < 0$ and volume expansion at 
$\eta > 0$ relative to the unstrained reference system. Meanwhile, the zinc-blende structure, in which there 
is only one type of bond length, Zn--O1 (Fig.\ref{fig5}b), presents a quite typical deformation behavior
when biaxially strained in the (001) plane. 

Figure~\ref{fig5}d shows the band gaps of (001)--oriented ZnO thin films as a function of biaxial strain for the 
two crystal structures. In the absence of any strain, we estimate an $E_{g}$ of $2.7$~eV for the wurtzite phase 
and $2.4$~eV for the zinc-blende phase. These results, in contrast to the systems analyzed previously, are not in 
good agreement with the experimental value of $3.3$~eV \cite{choi18}. A possible explanation for such a large band 
gap discrepancy may be the very low density of electronic states found at the bottom of the CB (Fig.\ref{fig5}e), 
which may complicate the estimation of $E_{g}$ both at the experimental and theoretical levels. 

We find that the impact of $\eta$ on the band gap of ZnO thin films is practically negligible (at least in comparison 
to the CeO$_{2}$ and TiO$_{2}$ cases), which is consistent with previous experimental observations \cite{choi18}. For 
instance, according to our first-principles calculations, $E_{g}$ for the wurtzite phase increases by just $2$\% at 
$\eta = +5$\% and decreases by $3$\% at $\eta = -5$\%; in comparison, a reduction of $4$\% was obtained at $\eta \approx 
-5$\% in experimental work \cite{choi18}. Meanwhile, the band gap for the zinc-blende phase is changed by even smaller
amounts, with only a $0.5$\% reduction at $\eta = +7$\% and a $0.5$\% increase at $\eta = -7$\% (note the opposite sign 
in the strain-driven $E_{g}$ change as compared to the wurtzite case, Fig.\ref{fig5}d). We note that the band gap of ZnO 
remains direct for all $\eta$ (Supplementary Figure~7). 

The reason behind the marginal effects of biaxial strain on the $E_{g}$ of ZnO thin films seems to be related to the fact 
that the majority of electronic states forming both the top of the VB and the bottom of the CB are oxygen $2p$ orbitals 
(Figs.\ref{fig5}e-f and Supplementary Figure~8). As a consequence of such an electronic band-structure symmetry the VB 
and CB edges are shifted very similarly when the crystal is subjected to external stress, thus leaving $E_{g}$ practically 
invariant.

\begin{table*}
\centering
\begin{tabular}{c c c c c c c c c}
\hline
\hline
$ $ & $ $ & $ $ & $ $ & $ $  \\
	$ $ & \multicolumn{2}{c}{\qquad \qquad $\frac{\Delta \chi}{\chi}$ \qquad \qquad} & \multicolumn{2}{c}{\qquad \qquad $\frac{\Delta d}{d}$ \qquad \qquad \qquad} & \multicolumn{2}{c}{\qquad $\frac{\Delta E_{\rm g}}{E_{\rm g}}~(\%)[{\rm model}]$ \qquad} & \multicolumn{2}{c}{\qquad $\frac{\Delta E_{\rm g}}{E_{\rm g}}~(\%)[{\rm DFT}]$ \qquad} \\
$ $ & $ $ & $ $ & $ $ & $ $  \\
\hline
$ $ & $ $ & $ $ & $ $ & $ $ & $ $ & $ $ & $ $ & $ $ \\
$\qquad \qquad \eta \qquad \qquad $ & $ -7\% $ & $ +7\% $ & $ -7\% $ & $ +7\% $ & $ -7\% $ & $ +7\% $ & $ -7\% $ & $ +7\% $ \\
$ $ & $ $ & $ $ & $ $ & $ $ & $ $ & $ $ & $ $ & $ $ \\
\hline
$ $ & $ $ & $ $ & $ $ & $ $ & $ $ & $ $ & $ $ & $ $ \\
	${\rm CeO_{2}~(111)}$ & $-0.063 $ & $-0.043 $ & $-0.014 $ & $0.025 $ & $+5 $ & $-1  $ & $+10 $ & $-4  $ \\
$ $ & $ $ & $ $ & $ $ & $ $ & $ $ & $ $ & $ $ & $ $ \\
	${\rm CeO_{2}~(001)}$ & $0.033  $ & $0.136  $ & $-0.029 $ & $0.038 $ & $+1  $ & $-11 $ & $+2  $ & $-29 $ \\
$ $ & $ $ & $ $ & $ $ & $ $ & $ $ & $ $ & $ $ & $ $ \\
	${\rm (a)~TiO_{2}~(001)}$ & $-0.109 $ & $0.063  $ & $-0.008 $ & $0.027 $ & $+6 $ & $-6 $ & $+21 $ & $-14 $ \\
$ $ & $ $ & $ $ & $ $ & $ $ & $ $ & $ $ & $ $ & $ $ \\
	${\rm (w)~ZnO~(001)}    $ & $0.022  $ & $-0.039  $ & $-0.020 $ & $0.026 $ & $+1 $ & $-1  $ & $-4  $ & $+2  $ \\
$ $ & $ $ & $ $ & $ $ & $ $ & $ $ & $ $ & $ $ & $ $ \\
	${\rm (zb)~ZnO~(001)}   $ & $0.074  $ & $ 0.061  $ & $-0.010 $ & $0.012 $ & $-3 $ & $-4  $ & $+1  $ & $-1 $ \\
$ $ & $ $ & $ $ & $ $ & $ $ & $ $ & $ $ & $ $ & $ $ \\
\hline
\hline
\end{tabular}
\label{tab:summary}
\caption{Relative band gap changes calculated at the maximum compressive and tensile biaxial stresses considered in this
	study for different semiconductor binary oxides. $\Delta \chi$ and $\Delta d$ stand for the variation in the  
	dielectric susceptibility and average metal-oxygen bond length as referred to the unstrained case. $(a)$, $(w)$, 
	and $(zb)$ stand for anatase, wurtzite, and zinc-blende structures, respectively. ${\rm [model]}$ refers to the 
	relative band gap variations estimated with the analytical model expressed in Eq.(\ref{eq:model}), while ${\rm [DFT]}$ 
	to those values obtained directly from the DFT calculations.}
\end{table*}

\subsection{Explanation of the observed $E_{g}$ trends using a simple model}
\label{subsec:model}
In previous sections we have demonstrated by means of computational first-principles methods that biaxial strain can be 
used to alter substantially the optoelectronic and photocatalytic properties of some binary oxide semiconductors. However, 
the influence of $\eta$ on the electronic band structure properties of binary oxide materials appears to be non-systematic 
and consequently difficult to predict. For instance, compressive biaxial strain hardly has any effect on the band gap of 
(001)~CeO$_{2}$ thin films whereas it induces a large $E_{g}$ increase in both (111)~CeO$_{2}$ and (001)~TiO$_{2}$ 
(Secs.\ref{subsec:ceo2-111}--\ref{subsec:tio2-001}). On the other hand, the effects of $\eta$ on the structural properties 
of binary oxides (provided that are stable and do not undergo phase transitions) are quite regular and relatively easy to 
foresee. Essentially, the metal-oxygen bond lengths for all the analyzed oxides are stretched by an average value of $\sim 
1$\% under moderate tensile biaxial strains and reduced by roughly the same amount under compressive biaxial strains. 
These outcomes suggest that the origins of the optoelectronic and photocatalytic variations induced by biaxial strain 
cannot be explained uniquely in terms of simple structural changes \cite{yin10,zhou14,wagner02}. In fact, a more sophisticated 
and general understanding of how photocatalytic activity can be tuned through $\eta$ is highly desirable for improving the 
design and computational screening of potential energy materials.

According to well-established theories \cite{omar75}, the band gap of a material is influenced by the first Fourier
coefficient of the crystal field, which results from the superposition of atomic potentials within the solid. The 
crystal field essentially depends on the interatomic distances, dielectric screening (or, alternatively, quantity 
of charge in the atomic environment), and density of atoms in the solid \cite{sun01}. By shortening the atomic 
distances and/or weakening the dielectric screening, the crystal field is enhanced and as a result the band gap is 
widened. Conversely, by increasing the bond lengths and/or enhancing the dielectric screening the crystal field is 
depleted and the band gap is reduced. Therefore, there are two possible ways of modifying the crystal field along 
with the band gap in a structurally stable solid -- changing its lattice parameters and/or its dielectric properties 
\cite{sun01}.   

By combining deformation potential theory ($E_{g} (\sigma) \propto \sigma$, where $\sigma$ represents mechanical stress) 
\cite{yin10,wei99} and the Kramers-Kroning relation ($E_{g} (\chi) \propto \chi^{-1/2}$, where $\chi$ represents the 
dielectric susceptibility of the material) \cite{sun01,lu15}, we have deduced a simple analytical model that is able 
to describe qualitatively the band gap variations induced by biaxial strain on binary oxide semiconductors. Specifically, 
we express the $\eta$-driven relative variations of the band gap as:
\begin{equation}
	\frac{\Delta E_{g}(\eta)}{E_{g}(0)} = -\frac{1}{2}\left( \frac{\Delta \chi (\eta)}{\chi (0)} + 2 
	\cdot \frac{\Delta d (\eta)}{d (0)} \right)~, 
\label{eq:model}
\end{equation}
where the dependence of each term on biaxial strain is explicitly noted (for instance, ``(0)'' denotes unstrained), 
$\Delta A (\eta) \equiv A(\eta) - A(0)$, $d$ represents the average metal-oxygen bond length, and the stress-strain 
relationship $\Delta \sigma / \sigma \approx - \Delta d / d$ has been used (the minus sign accounts for the fact that 
negative biaxial strains correspond to positive stresses and \emph{vice versa}). Our model treats the effects of 
structural and dielectric fluctuations on the band gap as independent quantities, that is, $E_{g} (\sigma,\chi) = 
f(\sigma) \cdot g(\chi)$ where $f$ and $g$ are the two functions specified above; because of this, the model cannot 
be quantitatively exact. Nevertheless, in spite of its limitations, this model provides a general and physically 
intuitive understanding of the results reported in the previous sections for binary oxides without the need to 
consider the details of the specific band structure of each material.   

Table~I shows the $\Delta E_{g} / E_{g}$ values obtained for biaxially strained CeO$_{2}$, TiO$_{2}$, and ZnO 
thin films at the largest $\eta$'s considered in this study using both DFT methods (exact) and the analytical model 
expressed in Eq.(\ref{eq:model}) (approximate). The corresponding $\Delta \chi / \chi$ and $\Delta d / d$ values 
are also reported in Table~I. The dielectric susceptibility is calculated with the formula $\chi = \epsilon_{\infty} 
-1$, where $\epsilon_{\infty}$ represents the ion-clamped dielectric constant of the material (estimated with 
perturbation DFT techniques) \cite{cazorla15}. In the case of CeO$_{2}$ and TiO$_{2}$ thin films Eq.(\ref{eq:model}) 
reproduces the relative band gap variations induced by biaxial strain correctly (albeit only at the qualitative 
level). For example, the analytical model predicts a reduction (increase) in $E_{g}$ under tensile (compressive) 
strain and the largest band gap variation is found for (001)--oriented CeO$_{2}$ thin films at $\eta > 0$. On the
other hand, the $\Delta E_{g} / E_{g}$ values estimated with Eq.(\ref{eq:model}) are systematically lower in
magnitude than those calculated directly with DFT techniques by roughly a factor of $2$. In the case of ZnO thin 
films, the analytical model consistently predicts band gap variations similar in magnitude to those obtained with
DFT methods; however, the corresponding signs are reversed in most cases. This comparison indicates that the 
limitations of Eq.(\ref{eq:model}) become more pronounced when trying to reproduce small $\Delta E_{g} / E_{g}$ 
values (as expected, given the qualitative nature of the model). 

The data shown in Table~I along with the analytical model in Eq.(\ref{eq:model}) allow us to understand
better the origins of the non-systematic $E_{g}$ trends found in binary oxides under biaxial strain.
As mentioned earlier, the $\Delta d / d$ changes induced by $\eta$ are systematic and general to most 
materials, but the $\Delta \chi / \chi$ variations are not. For example, for (001)--oriented CeO$_{2}$ 
thin films, the dielectric susceptibility always increases with increasing magnitude of biaxial strain,
irrespective of whether it is tensile or compressive. In contrast, for (001)--oriented TiO$_{2}$ anatase 
thin films, compressive (tensile) strain produces a reduction (increase) in $\chi$. Consequently, according 
to Eq.(\ref{eq:model}) and the data in Table~I, the changes induced by $\eta$ on the dielectric properties 
of each material are ultimately responsible for the irregular behavior observed in $\Delta E_{g} / E_{g}$. 
For instance, the large band gap changes found for (001)--oriented CeO$_{2}$ thin films under $\eta > 0$ and 
for (001)--oriented TiO$_{2}$ anatase thin films under $\eta < 0$ can be traced to the $\Delta \chi / \chi$ and 
$\Delta d / d$ terms having the same sign and hence being additive (Table~I). Likewise, the small band gap 
changes found for (111)--oriented CeO$_{2}$ thin films under $\eta > 0$ and for wurtzite (001)--oriented 
ZnO thin films either under $\eta > 0$ or $\eta < 0$ can be traced to the $\Delta \chi / \chi$ and $\Delta d 
/ d$ terms having opposite signs (Table~I). The new insights presented in this section in terms of the role
of dielectric susceptibility changes can be useful for better understanding the electronic behavior of binary 
oxide semiconductors subjected to biaxial strain, and for improving the design of potential photocatalytic 
materials.

\section{Conclusions}
\label{sec:conclusions}
In this study we have demonstrated by means of first-principles methods that biaxial strain can be used 
to tune the optoelectronic and photocatalytic properties of binary oxide semiconductors in a substantial and
predictable manner. The changes caused by $\eta$ on the band gap of simple oxides can be understood
in terms of structural and dielectric susceptibility variations that alter the crystal field and thus the 
bonding--antibonding splitting. The structural changes induced by biaxial strain on binary oxides are quite 
general and regular, but the accompanying changes in electronic screening are not systematic. As a result of 
those two distinct and sometimes opposing contributions, we find that the biaxial strain has different effects
on the band gap of different oxides -- the band gaps of stoichiometric CeO$_{2}$ and anatase TiO$_{2}$ can be 
reduced below $3$~eV under moderate tensile strains of $\approx +2$\% and $+3$\%, respectively, whereas the 
band gap of ZnO is only marginally affected (namely, $|\Delta E_{g}| / E_{g} \sim 1$\%) by $|\eta|$'s as large 
as $7$\%. We have also shown that $\eta$ has a significant influence on the band gap of non-stoichiometric and 
metal-doped binary oxide semiconductors. Furthermore, biaxial strain can alter significantly the energy levels 
of the valence and conduction band edges. In particular, we find that under tensile strain $(111)$--oriented 
CeO$_{2}$ and $(001)$--oriented TiO$_{2}$ thin films become suitable photocatalysts for driving the splitting 
of H$_{2}$O into H$_{2}$ and O$_{2}$ and the reduction of CO$_{2}$ into CH$_{4}$, respectively, in aqueous 
environment under sunlight. Strain engineering, therefore, offers a systematic design tool for achieving 
improved photocatalytic activity in binary oxide semiconductors, either on its own or in combination with other 
nanostructuring approaches.

\section*{ACKNOWLEDGMENTS}
Computational resources and technical assistance were provided by the Australian Government
and the Government of Western Australia through the National Computational Infrastructure
(NCI) and Magnus under the National Computational Merit Allocation Scheme and The Pawsey
Supercomputing Centre.


\begin{thebibliography}{30}
\bibitem{fajrina19} Fajrina, N. $\&$ Tahir, M.
	            A critical review in strategies to improve photocatalytic water splitting towards hydrogen production.
		    \textit{Int. J. Hydrog. Energy} \textbf{44}, 540 (2019).

\bibitem{xie16} Xie, S., Zhang, Q., Liu, G. $\&$ Wang, Y.
	        Photocatalytic and photoelectrocatalytic reduction of CO$_{2}$ using heterogeneous catalysts with controlled
		nanostructures.
		\textit{Chem. Commun.} \textbf{52}, 35 (2016).

\bibitem{shenoy19} Shenoy, J., Hart, J. N., Grau-Crespo, R., Allan, N. L. $\&$ Cazorla, C.
	           Mixing Thermodynamics and Photocatalytic Properties of GaP-ZnS solid solutions.
		   \textit{Adv. Theory Simul.} \textbf{2}, 1800146 (2019).

\bibitem{park16} Park, H., Kim, H.-I., Moon, G.-H. $\&$ Wonyong, C. 
	         Photoinduced charge transfer processes in solar photocatalysis based on modified TiO$_{2}$.
		 \textit{Energy Environ. Sci.} \textbf{9}, 411 (2016).

\bibitem{rahimi16} Rahimi, N., Pax, R. A. $\&$ Gray, E. M.
	           Review of functional titanium oxides.~I: TiO$_{2}$ and its modifications. 
		   \textit{Prog. Solid State Ch.} \textbf{44}, 86 (2016).

\bibitem{castano15} Castano, C. E., O'Keefe, M. J. $\&$ Farenholtz, W. G. 
	            Cerium-based oxide coatings.
		    \textit{Curr. Opin. Solid St. M.} \textbf{19}, 69 (2015).

\bibitem{montini16} Montini, T., Melchionna, M., Monai, M. $\&$ Fornasiero, P.
	            Fundamentals and catalytic applications of CeO$_{2}$‐-based materials.
		    \textit{Chem. Rev.} \textbf{116}, 5987 (2016).

\bibitem{kowalsa08} Kowalsa, E., Remita, H., Colbeau-Justin, C., Hupka, J. $\&$ Belloni, J.
	            Modification of titanium dioxide with platinum ions and clusters:~Application in photocatalysis.
                    \textit{J. Phys. Chem. C} \textbf{112}, 1124 (2008). 
	    
\bibitem{asahi01} Asahi, R., Morikawa, T., Ohwaki, T., Aoki, K. $\&$ Taga, Y.
	          Visible-light photocatalysis in nitrogen-doped titanium oxides.
		  \textit{Science} \textbf{293}, 269 (2001).

\bibitem{mitsudome11} Mitsudome, T., Mikami, Y., Matoba, M., Mizugaki, T., Jitsukawa, K. $\&$ 
	              Kaneda, K.
		      Design of a silver-cerium dioxide core-shell nanocomposite catalyst for 
		      chemoselective reduction reactions.
		      \textit{Angew Chem. Int. Ed.} \textbf{51}, 136 (2012).

\bibitem{takanabe17} Takanabe, K.
	             Photocatalytic water splitting: Quantitative approaches toward photocatalyst by design.
                     \textit{ACS Catal.} \textbf{7}, 8006 (2017).

\bibitem{hu18} Hu, S., Cazorla, C., Xiang, F., Ma, H., Wang, J., Wang, J., Wang, X., Ulrich, C., Chen, L. $\&$ Seidel, J.
	       Strain control of giant magnetic anisotropy in metallic perovskite SrCoO$_{3-\delta}$ thin films.
	       \textit{ACS Appl. Mater. Interfaces} \textbf{10}, 22348 (2018).	

\bibitem{heo17} Heo, Y., Hu, S., Sharma, P., Kim, K.-E., Jang, B.-K, Cazorla, C., Yang, C.-H., $\&$ Seidel, J.
	        Impact of isovalent and aliovalent doping on mechanical properties of mixed phase BiFeO$_{3}$.
		\textit{ACS Nano} \textbf{11}, 2805 (2017).

\bibitem{gopal17} Gopal, C. B. \emph{et al.}
	          Equilibrium oxygen storage capacity of ultrathin CeO$2-\delta$ depends non-monotonically
		  on large biaxial strain.
		  \textit{Nat. Commun.} \textbf{8}, 15360 (2017).

\bibitem{benson17} Benson, E. E. \emph{et al.}
                   Semiconductor-to-metal transition in rutile TiO$_{2}$ induced by tensile strain.
		   \textit{Chem. Mater.} \textbf{29}, 2173 (2017).

\bibitem{choi18} Choi, H.-J., Jang, W., Mohanty, B. C., Jung, Y.S., Soon, A. $\&$ Cho, Y.-S.
                 Origin of prestress-driven optical modulations of flexible ZnO thin films
                 processed in stretching mode.
                 \textit{J. Phys. Chem. Lett.} \textbf{9}, 5934 (2018).

\bibitem{yin10} Yin, W.-J., Che, S., Yang, J.-H., Gong, X.-G., Yan, Y. $\&$ Wei, S.-H.
	        Effective band gap narrowing of anatase TiO$_{2}$ by strain along a soft crystal direction.
		\textit{Appl. Phys. Lett.} \textbf{96}, 221901 (2010).

\bibitem{kelaidis18} Kelaidis, N., Kordatos, A., Christopoulos, S.-R.G. $\&$ Chroneos, A.
	             A roadmap of strain in doped anatase TiO$_{2}$.
		     \textit{Sci. Rep.} \textbf{8}, 12790 (2018).

\bibitem{zhou14} Zhou, W., Liu, Y., Yang, Y. $\&$ Wu, P.
	         Band gap engineering of SnO$_{2}$ by epitaxial strain: Experimental and theoretical
		 investigations.
		 \textit{J. Phys. Chem. C} \textbf{118}, 6448 (2014).

\bibitem{yan12} Yan, Q. M., Rinke, P., Winkelnkemper, M., Qteish, A., Bimberg, D., Scheffler, M. $\&$
                Van de Walle, C. G.
		Strain effects and band parameters in MgO, ZnO, and CdO.
		\textit{Appl. Phys. Lett.} \textbf{101}, 152105 (2012).

\bibitem{wagner02} Wagner, J.-M. $\&$ Bechstedt, F.
	           Properties of strained wurtzite GaN and AlN: Ab initio studies.
		   \textit{Phys. Rev. B} \textbf{66}, 115202 (2002).

\bibitem{cazorla15a} Cazorla, C.
	             The role of density functional theory methods in the prediction of nanostructured gas-adsorbent materials.
		     \textit{Coord. Chem. Rev.} \textbf{300}, 142 (2015).

\bibitem{cazorla12b} Cazorla, C., Rojas-Cervellera, V. $\&$ Rovira, C.
	             Calcium-based functionalization of carbon nanostructures for peptide immobilization in aqueous media.
		     \textit{J. Mater. Chem.} \textbf{22}, 19684 (2012).

\bibitem{cazorla17a} Cazorla, C. $\&$ Boronat, B.
	             Simulation and understanding of atomic and molecular quantum crystals.
		     \textit{Rev. Mod. Phys.} \textbf{89}, 035003 (2017).

\bibitem{pbesol} Perdew, J. P., Ruzsinszky, A., Csonka, G. I, Vydrov, O. A, Scuseria, G. E., Constantin, L. A., Zhou, X. $\&$
                 Burke, K.
                 Restoring the density-gradient expansion for exchange in solids and surfaces.
                 \textit{Phys. Rev. Lett.} \textbf{100}, 136406 (2008).

\bibitem{vasp} Kresse, G. $\&$ F\"urthmuller, J.
               Efficient iterative schemes for ab initio total-energy calculations using a plane-wave basis set.
               \textit{Phys. Rev. B} \textbf{54}, 11169 (1996);
               Kresse, G. $\&$ Joubert, D.
               From ultrasoft pseudopotentials to the projector augmented-wave method.
               \textit{Phys. Rev. B} \textbf{59}, 1758 (1999).

\bibitem{hubbard} Dudarev, S. L., Botton, G. A., Savrasov, S. Y., Humphreys, C. J. $\&$ Sutton, A. P.
	          Electron-energy-loss spectra and the structural stability of nickel oxide: An LSDA+U study.
                  \textit{Phys. Rev. B} \textbf{57}, 1505 (1998).

\bibitem{paw} Bl\"ochl P. E.
                  Projector augmented-wave method.
                  \textit{Phys. Rev. B} \textbf{50}, 17953 (1994).

\bibitem{kpoint} Monkhorst, H. J. $\&$ Pack, J. D.
	         Special points for Brillouin-zone integrations.
		 \textit{Phys. Rev. B} \textbf{13}, 5188 (1976).

\bibitem{cazorla15} Cazorla, C. $\&$ Stengel, M.
                    Electrostatic engineering of strained ferroelectric perovskites from first-principles.
                    \textit{Phys. Rev. B} \textbf{92}, 214108 (2015).

\bibitem{cazorla17} Cazorla, C.
                    Lattice effects on the formation of oxygen vacancies in perovskite thin films.
                    \textit{Phys. Rev. Appl.} \textbf{7}, 044025 (2017).

\bibitem{cazorla17b} Cazorla, C., Di\'eguez, O. $\&$ ${\rm \acute{I}}$${\rm \tilde{n}}$iguez, J.
                     Multiple structural transitions driven by spin-phonon couplings in a perovskite oxide.
                     \textit{Sci. Adv.} \textbf{3}, e1700288 (2017).

\bibitem{cazorla18} Cazorla, C. $\&$ \'I\~niguez J.
	            Giant direct and inverse electrocaloric effects in multiferroic thin films.
		    \textit{Phys. Rev. B} \textbf{98}, 174105 (2018).

\bibitem{hse06} Krukau, A. V., Vydrov, O. A., Izmaylov, A. F. $\&$ Scuseria, G. E.
                Influence of the exchange screening parameter on the performance of screened hybrid functionals.
                \textit{J. Chem. Phys.} \textbf{125}, 224106 (2006).

\bibitem{moses11} Moses, P. G., Miao, M., Yan, Q., $\&$ Van de Walle, C. G.
	          Hybrid functional investigations of band gaps and band alignments for AlN, GaN, InN, and InGaN.
		  \textit{J. Chem. Phys.} \textbf{134}, 084703 (2011).

\bibitem{resta88} Baldereschi, A., Baroni, S. $\&$ Resta, R. 
	          Band offsets in lattice-matched heterojunctions: A model and first-principles calculations for GaAs/AlAs.
		  \textit{Phys. Rev. Lett.} \textbf{61}, 734 (1988).

\bibitem{cazorla12} Cazorla, C. $\&$ Stengel, M.
	            First-principles modeling of Pt/LaAlO$_{3}$/SrTiO$_{3}$ capacitors under an external bias potential.
		    \textit{Phys. Rev. B} \textbf{85}, 075426 (2012).

\bibitem{goubin04} Goubin, F., Rocquefelte, X., Whangbo, M.-H., Montardi, Y., Brec, R. $\&$ Jobic, S.
	           Experimental and theoretical characterization of the optical properties of CeO$_{2}$, SrCeO$_{3}$, 
	           and Sr$_{2}$CeO$_{4}$ containing Ce$^{4+}$ ($f^{0}$) ions.
	           \textit{Chem. Mater.} \textbf{16}, 662 (2004).

\bibitem{wen18} Wen, X.-J., Niu, C.-G., Zhang, L., Liang, C. $\&$ Zeng, G.-M.
	        A novel Ag$_{2}$O/CeO$_{2}$ heterojunction photocatalysts for photocatalytic degradation 
		of enrofloxacin: possible degradation pathways, mineralization activity and an in depth 
		mechanism insight.
		\textit{Appl. Catal. B-Environ.} \textbf{221}, 701 (2018).

\bibitem{wei99} Wei, S.-H. $\&$ Zunger, A.
	        Predicted band-gap pressure coefficients of all diamond and zinc-blende semiconductors:
		Chemical trends.
		\textit{Phys. Rev. B} \textbf{60}, 5404 (1999).

\bibitem{mofarah19} Mofarah, S. S., Adabifiroozjaei, E., Webster, R., Koshy, P., Nekouei, R. K., Lim, S.,
	            Yao, Y., Cazorla, C., Liu, Z., Lambropoulos, N. $\&$ Sorrell, C. C.
		    Proton-assisted creation of controllable volumetric oxygen vacancy in ultrathin ceria 
		    for pseudo-capacitive energy storage applications.
		    \textit{Nat. Commun.} \textbf{10}, 2594 (2019).

\bibitem{korobko12} Korobko, R., Patlolla, A., Kossoy, A., Wachtel, E., Tuller, H. L., Frenkel, A. I. 
	            $\&$ Lubomirsky, I.
		    Giant electrostriction in Gd-Doped ceria.
		    \textit{Adv. Mater.} \textbf{24}, 5857 (2012).

\bibitem{yamada12} Yamada, Y. $\&$ Kanemitsu, Y.
	           Determination of electron and hole lifetimes of rutile and anatase TiO$_{2}$ single crystals.
		   \textit{Appl. Phys. Lett.} \textbf{101}, 133907 (2012).

\bibitem{batzill11} Batzill, M.
	            Fundamental aspects of surface engineering of transition metal oxide photocatalysts.
		    \textit{Energy Environ. Sci.} \textbf{4}, 3275 (2011).

\bibitem{choi94} Choi, W., Termin, A. $\&$ Hoffmann, M. R. 
	         The role of metal ion dopants in quantum-sized TiO$_{2}$: Correlation between photoreactivity 
		 and charge carrier recombination dynamics.
		 \textit{J. Phys. Chem. C} \textbf{98}, 13669 (1994).

\bibitem{bousquet10} Bousquet, E., Spaldin, N. A. $\&$ Ghosez, P.
	             Strain-induced ferroelectricity in simple rocksalt binary oxides.
		     \textit{Phys. Rev. Lett.} \textbf{104}, 037601 (2010).

\bibitem{ozgur05} Ozgur, U., Alivov, Y. I., Liu, C., Teke, A., Reshchikov, M. A., Dogan, S., Avrutin, V., 
		  Cho, S.-J. $\&$ Morkoc, H. 
		  A comprehensive review of ZnO materials and devices.
		  \textit{J. Appl. Phys.} \textbf{98}, 041301 (2005).

\bibitem{wang13} Wang, Z., Wang, F., Wang, L., Jia, Y. $\&$ Sun, Q.
	         First-principles study of negative thermal expansion in zinc oxide.
		 \textit{J. Appl. Phys.} \textbf{114}, 063508 (2013).

\bibitem{omar75} Omar, M. A.
	         \textit{Elementary Solid State Physics: Principles and Applications}
                 (New York: Addison-Wesley, 1975).

\bibitem{sun01} Sun, C. Q., Sun, X. W., Tay, B. K., Lau, S. P., Huang, H. T. $\&$ Li, S.
	        Dielectric suppression and its effect on photoabsorption of nanometric semiconductors.
		\textit{J. Phys. D: Appl. Phys.} \textbf{34}, 2359 (2001).

\bibitem{lu15} Haiming, L. $\&$ Meng, X.
	       Correlation between band gap, dielectric constant, Young's modulus and melting 
	       temperature of GaN nanocrystals and their size and shape dependences.
	       \textit{Sci. Rep.} \textbf{5}, 16939 (2015).

\end{thebibliography}
\end{document}